\documentclass[twocolumn,twoside,slac,floatfix]{revtex4}
\usepackage{graphicx}
\usepackage{fancyhdr}
\begin{document}

\title{Event Data Definition in LHCb}

\author{Marco Cattaneo}
\affiliation{CERN, 1211 Geneva, Switzerland}
\author{Gloria Corti}
\affiliation{CERN, 1211 Geneva, Switzerland}
\author{Markus Frank}
\affiliation{CERN, 1211 Geneva, Switzerland}
\author{Pere Mato Vila}
\affiliation{CERN, 1211 Geneva, Switzerland}
\author{Silvia Miksch}
\affiliation{TU Vienna, 1040 Vienna, Austria}
\author{Stefan Roiser}
\affiliation{CERN, 1211 Geneva, Switzerland}

\begin{abstract}
We present the approach used for defining the event object model for the LHCb 
experiment. This approach is based on a high level modelling language, which is
independent of the programming language used in the current implementation of 
the event data processing software. The different possibilities of object 
modelling languages are evaluated, and the advantages of a dedicated model 
based on XML over other possible candidates are shown. After a description of 
the language itself, we explain the benefits obtained by applying this approach
in the description of the event model of an experiment such as LHCb. Examples 
of these benefits are uniform and coherent mapping of the object model to the 
implementation language across the experiment software development teams, easy 
maintenance of the event model, conformance to experiment coding rules, etc.\\

The description of the object model is parsed by means of a so called front-end
which allows to feed several back-ends. We give an introduction to the model 
itself and to the currently implemented back-ends which produce information 
like programming language specific implementations of event objects or meta 
information about these objects. Meta information can be used for introspection
of objects at run-time which is essential for functionalities like object 
persistency or interactive analysis.  This object introspection package for C++
has been adopted by the LCG project as the starting point for the LCG object 
dictionary that is going to be developed in common for the LHC experiments.\\

The current status of the event object modelling and its usage in LHCb are 
presented and the prospects of further developments are discussed.\\

\end{abstract}

\maketitle

\thispagestyle{fancy}

  \section {\label{SECT_INTR}INTRODUCTION}
  This paper gives an overview of tools which are used for the
  description and subsequent handling of event data objects in LHCb
  \cite{MOU} which is one of four experiments being prepared at the
  Large Hadron Collider machine (LHC) at the European Institute for
  High Energy Physics (CERN) due to begin operation in 2007. The work
  was carried out as part of Gaudi \cite{Mato:01a, Cattaneo:00a},
  which is the software framework for the LHCb experiment.\\

  The LHCb experiment is supposed run for at least ten
  years and the amount of data that will be stored is expected to be
  in the order of several Peta bytes.\\

  The work described in this paper concentrates on the
  modelling of the reconstructed data and the data retrieved after the
  analysis process. For the rest of the paper these two models will be
  referred to as the LHCb Event Model.\\

  In the next section (\ref{SECT_REQU}) the requirements and
  prerequisites for these description tools will be discussed. Section
  \ref{SECT_MODE} contains an in-depth discussion of the model that
  was developed for carrying out the tasks followed by section
  \ref{SECT_CHOI} which describes the different possibilities for the
  implementation of the the model and the choices which were  made. An
  example class will be shown in section \ref{SECT_EXAM}. Section
  \ref{SECT_EVAL} contains an evaluation of the model in respect of
  user acceptance and usability. Section \ref{SECT_IMPR} gives some
  details about the possible future improvements and and outlook. The
  paper will be concluded by a summary in section \ref{SECT_SUMM}. 

  \section {\label{SECT_REQU}REQUIREMENTS}
  The design of the LHCb Event Model was constrained by
  several requirements. These requirements arose both from the user
  and the technical side.\\  

  Requirements from the user side were such as:
  \begin{itemize}
  \item \textbf{Long Lifetime:} Including the construction and planning phase,
    the LHCb experiment is supposed to run for more than two decades. In this
    respect the durability of the described data is important. For example
    it should be always possible to read data back into a program that was
    created several years ago by a different program and with a different layout 
    of the data.
  \item \textbf{Language Independence:} As the experiment software
    will continue to evolve when the experiments are up and running,
    new languages are likely to come up which are more adequate for
    the software framework and with better functionality for the
    software developers. In order not to reimplement the
    event model every time a new language is introduced it would be
    important to describe the event model with a higher level language
    from which concrete implementations can be derived.
  \item \textbf{Easiness of Design:} Physicists describing event data should
    not be bothered with complex implementation languages which are
    difficult to learn and understand. The goal is to either create
    a language which is easy to understand and to learn with a simple
    syntax or a language potential users are already used to and so
    can use ad-hoc. 
  \item \textbf{Short descriptions:} Data descriptions in concrete
    implementations are often verbose and so error prone to
    implement. In C++ e.g. implementing a data member in a class also
    requires the implementation of a setter- and a getter-method and
    information of this member will also appear in many other places
    of an implementation of a class, e.g. the streaming functions for
    output.
  \end{itemize}
  On the other side there were also technical constraints such as:
  \begin{itemize}
  \item \textbf{Artificial Constraints:} As the LHCb software framework is
    written in object oriented style, the event model should also be
    capable of reflecting these concepts. But not all capabilities
    of current programming languages need to be reflected in such a
    description language. While concepts like bitfields would be useful
    to implement, other concepts like abstract interfaces are perhaps not
    necessary.
  \item \textbf{Modelling Relations:} There were also requirements
    data modelling that are not reflected in current object oriented
    programming languages directly, e.g. the distinction between
    data members which are holding some data of an object and
    relations which point to other parts of the event model. 
  \end{itemize}

  \begin{figure*}[!htb]
   \fbox{\includegraphics[width=.85\textwidth]{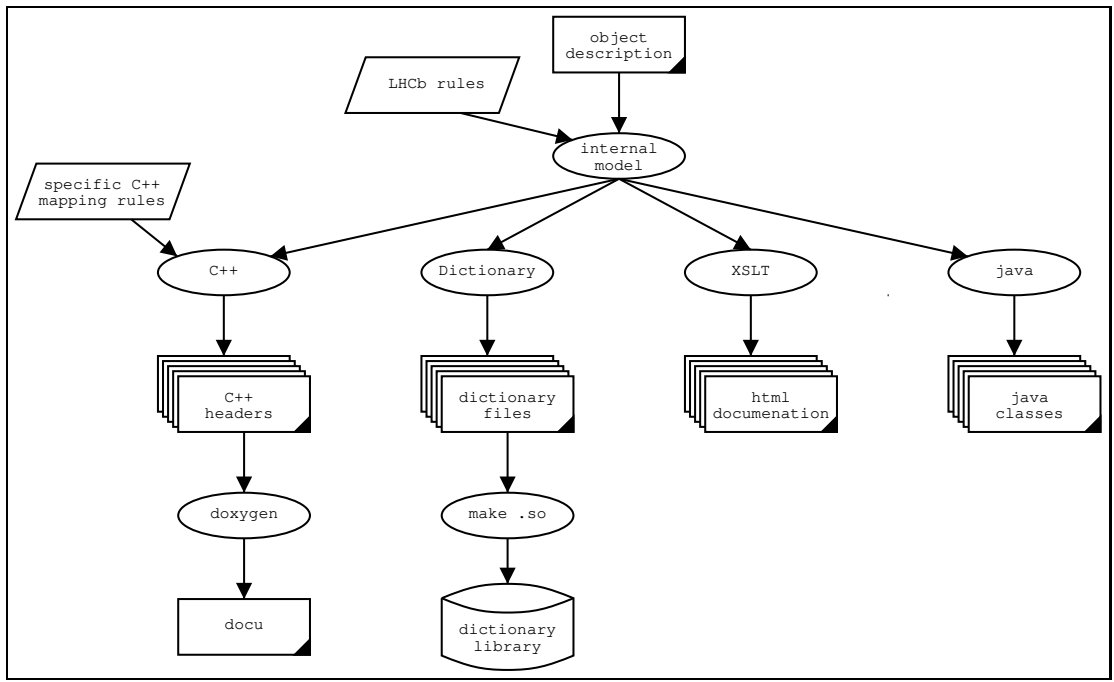}}
   \caption{\label{FIGU_OVER}Object Description Model Overview}
  \end{figure*}

  \section {\label{SECT_MODE}THE MODEL}
  To ensure that the requirements are met a model which
  describes the event data structure with the means of a
  high level language was designed. The usage of a high level language
  for the description of the data structures ensures flexibility and
  durability of the model.\\

    \subsection{Overall View}
    The overall design of the tools (see Figure \ref{FIGU_OVER}) was
    divided into two  parts. A front-end, which will parse the
    definition language and fill an in memory representation of the
    processed data, and back-ends which will use this
    memory representation to produce whatever output is needed. These
    back-ends will produce a representation of the data in a given
    implementation language (e.g. C++) but also other kinds of
    information such as a meta representation of the data, used for
    reflection purpose of the data (see section \ref{SSEC_META}). 

    \subsection{Front-end}
    The front-end will parse object descriptions written
    by the users. These object definitions are the only input to the
    system that the users have to maintain. After parsing these
    descriptions an in memory representation of the objects will be
    produced. The goal was to define a language that describes objects
    on an abstract level and does not need to be changed when new
    back-ends are implemented. With this technique the long lifetime of
    the object description will be guaranteed. 

    \subsection{Back-ends}
    There are a number of possible back-ends that have been developed. 

      \subsubsection{C++}
      As the current implementation of the Gaudi framework is
      currently done in C++, therefor the first goal was to implement a
      back-end which will produce a representation of the objects in
      C++. The capabilities of the C++ back-end are limited in the
      sense of C++, as the full functionality of C++ was not needed for
      representing the event model in this language. The
      main functionality needed was to represent members of objects
      and relations between them. The members are translated into
      class members of C++ classes. The relations are handled by
      an internal mechanism of the Gaudi framework \cite{Frank:00a}.\\

      The goal of this back-end is to produce C++ header files which
      contain the object descriptions of the event model. The back-end will
      also produce implementations of simple functions like accessors
      to members or serializing functions. Implementations of more
      complex functions will be left to the user. With the means of an
      internal database the back-end will also be capable of including
      most of the necessary header files for a given class.\\

      In Gaudi, like in many other large
      software frameworks, exist some coding guidelines. The goal of
      this back-end is not only to reflect the structure of the
      objects but also meet these guidelines. These coding guidelines
      guarantee a uniform layout and look-and-feel of the generated
      classes. This can be an advantage for people having to work
      with this code, as not only the style but also the structure of
      all classes generated by the back-end will be the same.\\ 

      In addition to the source code also some
      documentation for the classes and objects were generated. In a
      subsequent step it is though possible to extract this information and
      generate some general description about the event objects from
      it. The so retrieved documentation can either be viewed on
      webpages or printed in different formats.

  \begin{figure*}[!bht]
   \fbox{\includegraphics[angle=270,width=.85\textwidth]{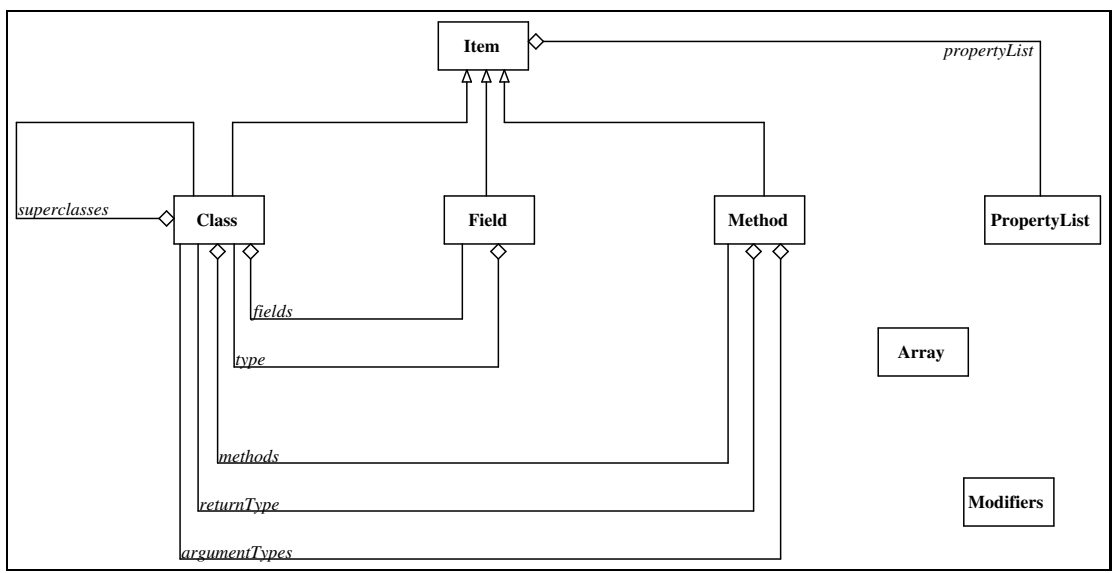}}
   \caption{\label{FIGU_REFL}Reflection Model}
  \end{figure*}

      \subsubsection{\label{SSEC_META}C++ Meta-Representation}
      Today many modern object oriented and scripting languages (e.g. Java, Python) 
      provide reflection capabilities on their objects. Reflection is the ability to query
      the internal structure of objects at runtime and also interact
      with them (i.e. set or get values of members or call
      functions). Reflection is essential for tasks like persistence of
      objects or interactive usage, for example when working with objects
      from a terminal prompt through a scripting language.\\ 

      A package for reflection on C++ (see Figure \ref{FIGU_REFL}) was developed in the
      Gaudi framework. The appropriate sources to fill this
      reflection package were also derived from the
      high level description of the objects. These descriptions are
      C++ classes which had the needed meta information about the
      objects and were compiled in a later step into libraries which
      then could be loaded by the reflection to provide the
      meta information about their objects.\\

      The reflection package itself is very simple and was derived from
      the java.lang.reflection package
      \footnote{see http://java.sun.com/j2se/1.4.1/docs/api\\/java/lang/reflect/package-summary.html}. 
      For the first implementation this
      model seems to be sufficient for the current needs but in later
      steps a redesign might be needed to better resemble the specific needs
      of the C++ language (e.g. pointers, references).

      \subsubsection{HTML}
      For documentation purpose the information about the
      objects can also be generated in a more human readable form, like
      HTML which then can be browsed with a a web browser.

  \section {\label{SECT_CHOI}TECHNICAL CHOICES}
  After defining the logical structure of the model, several decisions about the 
  concrete implementation had to be taken.

    \subsection{Description Language}
    The most important decision that had to be taken was the one about
    the description language itself which will be used for describing
    the event data. Several choices were possible:
    \begin{itemize}
    \item \textbf{Plain Text:} In previous experiments, such as of the 
      Large Electron Positron Collider (LEP) at CERN, this
      approach was used in some cases. The advantage of this approach
      would be the easiness of creation of such files, as no special
      environment, such as editors or compilers, would be needed. On the
      other hand the creators of such files could not have immediate
      feedback whether they created syntactically correct files until the
      data is read through the tool. Also the tool
      itself would be extremely difficult to implement as all the
      syntactic constraints needed to be invented.
    \item \textbf{C++:} C++ or any other object oriented
      implementation language should not be considered as the first choice
      for data description. As C++ is a complex language and
      difficult to maintain it should be the goal of the tools processing the
      data to produce output in this language rather than urging the users
      to provide description with it.
    \item \textbf{IDL:} IDL is used in many different environments such
      as Corba to describe interfaces. The syntax is very much restricted
      on C++ and so the possibility to use it as a durable language over
      several decades is very much limited as well.
    \item \textbf{UML:} UML again is a widely used language to describe
      data in the computing environment. This language, as IDL, has a very
      strict syntax and the possibilities for flexible extensions
      of this language are not optimal.
    \item \textbf{XML:} XML on the other hand is a very flexible language
      that also provides a very strict syntax, as this description of a
      syntax is already part of the language itself. The syntax of XML
      can either be described in so-called DTDs or XML Schemas. While
      DTD provides a limited functionality, XML Schema is a complex
      language which gives the developer of the syntax a lot of means
      to go to a very detailed level of description. XML is also a
      wide spread and well known language in the computing environment
      for which several tools such as browsers, editors, parsers and
      language bindings exist.\\

      The syntax of XML consists of two main entities, namely elements and
      attributes. Elements define the objects of the language while attributes
      are always parts of elements and specify their behaviour. An advantage 
      of XML is the possibility to specify default values for
      attributes. These default values can be specified in the syntax and if
      the attribute is not explicitly specified the default value will be taken.
      This is very convenient for users to shorten the descriptions and to
      save time when typing. As the developer of the language is also the creator
      of its syntax, extending it with new features is very simple and straightforward.
    \end{itemize}
    Because of its ability of easy extension and its
    strict syntax XML was chosen as the language for the description
    of the objects. It was also decided to start the description of
    the language syntax with a DTD and switch to XML Schema if the
    language reaches a level of complexity that DTD is not able to
    handle anymore.\\

    XML was also chosen because it was already used in LHCb for the detector 
    description. So it was hoped that people are already used to working with this 
    language and it will not take a lot of time for them to get up to speed with it.

    \subsection{Implementation Language}
    It was also necessary to decide on the
    implementation language for the tools which would be used for
    the parsing of the description language and the different back-ends
    of the system. In general any language that
    would be capable of parsing a given language and producing some
    output would be sufficient for this choice. Possible choices in this
    case were compiled object oriented languages such as C++ or Java as
    well as scripting languages like Python or Perl. Although the tool
    itself is completely independent of Gaudi, C++ was chosen for the
    implementation language because it is used throughout the
    framework.\\ 

    As a tool for parsing the description language,
    Xerces-C \footnote{see http://xml.apache.org/xerces-c/index.html}
    was chosen as there existed a C++ implementation of this
    parser and it was also already used in Gaudi for the
    detector description part. Xerces is also able to verify
    XML documents either with DTD or XML Schema.

  \section {\label{SECT_EXAM}EXAMPLE CLASS}
  To demonstrate parts of the capabilities of the system the current implementation
  of the MCParticle class was chosen. In Table \ref{TABL_MCPX} the XML
  description of the XML description is shown. Producing the C++
  header file out of this class will result in an file of 374 lines.

  \begin{table*}[t]
   \begin{ttfamily}
    \begin{center}
      \caption{\label{TABL_MCPX}MCParticle.xml}
      \begin{tabular}{|p{\textwidth}|}
	\hline
               $<$?xml version="1.0" encoding="UTF-8"?$>$\\
               $<$!DOCTYPE gdd SYSTEM "gdd.dtd"$>$\\
               $<$gdd$>$\\
\hspace{1.7mm} $<$package name="Event"$>$\\
\hspace{3.4mm} $<$class name="MCParticle" id="210" location="MC/Particles" author="G. Corti" desc="The..."$>$\\
\hspace{5.1mm} $<$base name="KeyedObject\&lt;int\&gt;"/$>$\\
\hspace{5.1mm} $<$attribute name="momentum" type="HepLorentzVector" init="0.0,0.0,0.0,0.0" desc="4-moment-..."/$>$\\
\hspace{5.1mm} $<$attribute name="particleID" type="ParticleID" init="0" desc="Particle ID" /$>$\\
\hspace{5.1mm} $<$attribute name="hasOscillated" type="bool" init="false" desc="Describe if a particle has..."/$>$\\
\hspace{5.1mm} $<$attribute name="helicity" type="double" desc="Helicity" /$>$\\
\hspace{5.1mm} $<$method name="virtualMass" type="double" const="TRUE" desc="Retrieve virtual mass"$>$\\
\hspace{6.8mm} $<$code$>$  return m$\_$momentum.m(); $<$/code$>$\\
\hspace{5.1mm} $<$/method$>$\\
\hspace{5.1mm} $<$method name="pt" type="double" const="TRUE" desc="Short-cut to pt value"$>$\\
\hspace{6.8mm} $<$code$>$ return m$\_$momentum.perp(); $<$/code$>$\\
\hspace{5.1mm} $<$/method$>$\\
\hspace{5.1mm} $<$method name="mother" type="const MCParticle*" const="TRUE" desc="Pointer to parent particle"$>$\\
\hspace{6.8mm} $<$code$>$ if( originVertex() ) { return originVertex()-$>$mother(); } else { return 0; }  $<$/code$>$\\
\hspace{5.1mm} $<$/method$>$\\
\hspace{5.1mm} $<$relation name="originVertex" type="MCVertex" desc="Pointer to origin vertex"/$>$\\
\hspace{5.1mm} $<$relation name="endVertices" type="MCVertex" multiplicity="M" desc="Vector of pointers to..."/$>$\\
\hspace{5.1mm} $<$relation name="collision" type="Collision" desc="Ptr to Collision to which the vertex be..."/$>$\\
\hspace{3.4mm} $<$/class$>$\\
\hspace{1.7mm} $<$/package$>$\\
               $<$/gdd$>$\\ 
     \hline
     \end{tabular}
    \end{center}
   \end{ttfamily}
  \end{table*}


  \section {\label{SECT_EVAL}EVALUATION}

    \subsection{XML}
    Although of having the drawback to be a verbose language it
    turned out that XML was a good choice for the
    description language. It allowed to start with a minimal
    functionality and enhance the language in very short
    development cycles when new functionality was requested by the
    user community. In fact the enhancing of the language was needed
    several times, so for example bitfields were introduced or some
    more detailed way to describe arguments of functions.

    \subsection{Acceptance by users}
    Several talks about the object description tools and its usage
    were given in meetings of the collaboration. Additionally a
    webpage with frequently asked questions was kept up to
    date. These actions led to a quite good acceptance by the user
    community and also speeded up the development cycles of the tool. 

    \subsection{Input-Output Ratio}
    The ratio between input and output code is calculated
    on the basis of lines of XML code and its generated C++ code. The
    input-output ratio of XML code to generated C++ source code is
    around 1:4. The overall ratio from XML code to all generated C++
    code is approximately 1:12. 

    \subsection{Usage so far}
    The usage of the object description tools by the users
    in LHCb started in December 2001. Since that time 24 iterations
    of the LHCb event model were produced. This seems to be a
    quite high number, but has to be seen in connection to the fact
    that the start of the usage of the tools was also the start of the
    redesign of the LHCb event model which was an urgent task at that
    time. 

  \section {\label{SECT_IMPR}FUTURE IMPROVEMENTS AND OUTLOOK}
  The software for object description was developed with the long
  lifetime of the experiment in mind. From this point of view the
  flexibility and extensibility of the software was a major
  concern. Extensions in the following fields can be carried
  out.

  \begin{itemize}
    \item \textbf{Extensions to the Language:}
    If needed new concepts for the object description language itself
    will be introduced. In principle there are three steps that need
    to be carried out. The syntax has to be changed, the front-end
    made aware of the new concept and finally the back-ends need
    retrieve the new information and produce the corresponding output.
    During the development phase of the package it
    was already proven that extending the language and the depending
    software is feasible in quite short development cycles which allow
    flexible adaptation to upcoming needs of the user community.

    \item \textbf{New Back-ends:}
    Not only changes to the language itself but also new back-ends
    could be needed in the future for e.g. C\# or other languages that may
    become important. In that case a new tool will be created. It will
    make use of the already existing front-end and the model that is
    filled with it. Walking through this model it will output the
    descriptions of the event model in the syntax of the new
    language. As language independence was a key issue when designing
    the software the new languages should be able to be filled with
    the existing syntax of the description language.

    \item \textbf{Integration with LCG software:}
    The LHC Computing Grid (LCG) is a new project at CERN which aims to provide 
    hard- and software computing facilities for the 4 upcoming experiments. Concerning
    the LCG software there are already some projects
    \cite{Duellmann:03a, Frank:03a, Mato:03a}. In the future LHCb will
    adopt these software packages and integrate them into the Gaudi framework. 

    \item \textbf{Improvements to Reflection:}
    The current implementation of  the reflection package was derived from the 
    java.lang.reflect package. The structure of this module is quite simple and was 
    appropriate for a first implementation of the package. Nevertheless C++ has some
    concepts which have no equivalence in Java, like pointers or references. In a later
    step a redesign of the reflection package might be envisaged, which will also
    lead to some adaptations for the generated code to fill the
    reflection module.

    \end{itemize}

  \section {\label{SECT_SUMM}SUMMARY}
  In this paper we introduced the concept of a high level language for
  description of concrete implementation languages of the LHCb Event
  Model. The key issues of the model like long lifetime, flexibility
  and durability were pointed out. The model itself was described in
  depth with its possibilities for future extensions. A concept for
  reflection in C++ was introduced which goes hand in hand with the
  object description tools which are able to fill it. It was also
  shown that the model has proven its usability for during more than a
  year and was accepted by the user community.


  

\end{document}